\documentclass[a4paper,10pt]{article}
\usepackage[utf8x]{inputenc}

\usepackage{microtype}%if unwanted, comment out or use option "draft"
\usepackage{amsmath}
\usepackage{amssymb}
\usepackage{url}
\usepackage{slashbox}
\usepackage[ruled, linesnumbered, vlined]{algorithm2e}

\usepackage{hyperref}
\hypersetup{
  colorlinks=true,
  citecolor=blue,
  linkcolor=blue,
  urlcolor=blue,
  bookmarksnumbered,
  bookmarksopen,
  pdftitle={Engineering Time-dependent One-To-All Computation},
  pdfauthor={Robert Geisberger},
  pdfsubject={Engineering Time-dependent One-To-All Computation},
  pdfkeywords={graph, priority queue, Dijkstra, routing, hierarchy, hierarchical routing, contraction hierarchy, ch, time-dependent, TCH},
}

\newcommand{\set}[1]{\left\{ #1\right\}}

\newcommand{\real}{\mathbb{R}}

\newcommand{\realrange}[2]{\left[#1, #2\right]}

\newcommand{\unitrange}[2]{\realrange{0}{1}}

\newcommand{\Oh}[1]{\mathrm{O}\!\left( #1\right)}

\newcommand{\llabel}[1]{\label{\labelprefix:#1}}
\newcommand{\labelprefix}{} % later redefined using renewcommand

\newcommand{\discussionsize}{\small}

\newcommand{\notiz}[1]{}
\newcommand{\frage}[1]{}
\newenvironment{code}{\noindent%\sf%
\begin{tabbing}%
\hspace{2em}\=\hspace{2em}\=\hspace{2em}\=\hspace{2em}\=\hspace{2em}\=%
\hspace{2em}\=\hspace{2em}\=\hspace{2em}\=\hspace{2em}\=\hspace{2em}\=%
\kill}{\end{tabbing}}

\newcommand{\labelcommand}{}
\newcommand{\captiontext}{}
\newsavebox{\codeparam}
\newcounter{lineNumber}
\newenvironment{disscodepos}[3]{%
\renewcommand{\labelcommand}{#2}%
\renewcommand{\captiontext}{#3}%
\sbox{\codeparam}{\parbox{\textwidth}{#3}}%
\begin{figure}[#1]\begin{center}\begin{code}\setcounter{lineNumber}{1}}{%
\end{code}\end{center}\caption{\llabel{\labelcommand}\captiontext}\end{figure}}

{\end{disscodepos}}

\newcommand{\Is}{\mbox{\rm := }}

\newdimen\endofsize\endofsize=0.5em

\renewcommand{\frage}[1]{[{\sf#1}]}

\newcommand{\myparagraph}[1]{\par\vspace{2mm}\noindent{\bf #1}}

\newcommand{\Gup}{G_{\uparrow}}
\newcommand{\Gdown}{G_{\downarrow}}

\newcommand{\Eup}{E_{\uparrow}}
\newcommand{\Edown}{E_{\downarrow}}

\newcommand{\fapprox}{f^\updownarrow}

\newcommand{\lb}[1]{#1^\downarrow} % lower bound
\newcommand{\ub}[1]{#1^\uparrow} % upper bound

\newcommand{\td}{\delta}
\newcommand{\link}[2]{#2*#1} % link two TTFs

\title{Engineering Time-dependent One-To-All Computation}
\author{Robert Geisberger\\
{\small Karlsruhe Institute of Technology, 76128 Karlsruhe, Germany}\\
{geisberger@kit.edu}
}

\begin{document}

\maketitle

\begin{abstract}
Very recently a new algorithm to the nonnegative single-source shortest path problem on road networks has been discovered.
It is very cache-efficient, but only on static road networks.
We show how to augment it to the time-dependent scenario.
The advantage if the new approach is that it settles nodes, even for a profile query, by scanning all downward edges.
We improve the scanning of the downward edges with techniques developed for time-dependent many-to-many computations.
\end{abstract}

\section{Introduction}

The new algorithm for the nonnegative single-source shortest path problem on road networks \cite{dgnw-phast-10} uses contraction hierarchies \cite{gssd-chfsh-08}.
First, it performs a forward upward search, and then it processes all nodes in descending order of importance.
The algorithm is so efficient because it can order the edges very cache-efficient and can be parallelized.
We consider the time-dependent scenario, where the travel time depends on the departure time, the edge weights are complex travel time functions (TTFs).
Such a TTF maps a departure time to the travel time.
We want to solve the problem of computing the travel time profiles from one source node to all other nodes in the graph.
A travel time profile is a TTF that maps the departure time at the source to the earliest arrival time at the target node.
Significantly the most work on solving our problem is to process the TTFs, the traversal of the graph takes negligible time.
So the cache-efficient order of the edges brings is no longer a significant speed-up.
But the way the edges are processed allows to prune a lot of expensive TTF operations by using approximate TTFs.

\subsection{Related Work}

Many new algorithms for the time-dependent point-to-point shortest path problem have been developed recently.
We refer to \cite{dw-tdrp-09} for an overview.
Also, a time-dependent version of contraction hierarchies (TCH) \cite{bdsv-tdch-09,bgns-tdcha-10} exists, and we use it for our new algorithm.
It was augmented to compute travel time tables by \cite{gs-etdmm-10}.

\section{Preliminaries}

% %%%%%%%%%%%%%%%%%%%%%%%%%%%%%%%%%%%%%%%%%%%%%%%%%%%%%%%%%%%%%%%%%%%%%%%%%%%%%
\subsection{Time-Dependent Road Networks}
% %%%%%%%%%%%%%%%%%%%%%%%%%%%%%%%%%%%%%%%%%%%%%%%%%%%%%%%%%%%%%%%%%%%%%%%%%%%%%
Let $G=(V,E)$ be a directed graph representing a road network.\footnote{Nodes represent junctions and edges represent road segments.}
Each edge $(u,v)\in E$ has a function $f:\real\to\real_{\ge 0}$ assigned as edge weight.
This function $f$ specifies the time $f(\tau)$ needed to reach $v$ from $u$ via edge $(u,v)$ when starting at \emph{departure time} $\tau$.
So the edge weights are called \emph{travel time functions} (TTFs).

In road networks we usually do not arrive earlier when we start later.
So all TTFs $f$ fulfill the \emph{FIFO-property\nocorr}:
$\forall\tau' > \tau: \tau' + f(\tau') \geq \tau + f(\tau)$.
In this work all TTFs are sequences of \emph{points} representing piecewise linear functions.\footnote{
	Here, all TTFs have period $\Pi = 24\mathrm{h}$.
	However, using non-periodic TTFs makes no real difference.
	Of course, covering  more than 24h will increase the memory usage.
}
%This enables us to represent the functions as finite sequences of points $\langle (x_1,y_1),\dots,(x_k,y_k) \rangle$ with $0 \leq x_1 < \dots < x_k<\Pi$.
With $|f|$ we denote the \emph{complexity} (i.e., the number of points) of $f$.
We define $f \sim g :\Leftrightarrow \forall \tau: f(\tau) \sim g(\tau)$ for $\sim \in \set{<,>,\le,\ge}$.

For TTFs we need the following three operations:
\begin{itemize}\itemsep0cm\topsep0cm
	\item \textit{Evaluation.}
		Given a TTF $f$ and a departure time $\tau$ we want to compute $f(\tau)$.
		Using a bucket structure this runs in constant average time.
	\item \textit{Linking.}
		Given two adjacent edges $(u,v),(v,w)$ with TTFs $f,g$ we want to compute the TTF of the whole path $\langle u \to_f v \to_g w\rangle$.
		This is the TTF $g*f:\tau\mapsto g(f(\tau)+\tau) + f(\tau)$ (meaning $g$ ``after'' $f$).
		It can be computed in $\Oh{|f|+|g|}$ time and $|g*f| \in \Oh{|f|+|g|}$ holds.
		Linking is an associative operation, i.e., $f*(g*h) = (f*g)*h$ for TTFs $f,g,h$.
	\item \textit{Minimum.}
		Given two parallel edges $e$, $e'$ from $u$ to $v$ with TTFs $f,f'$, we want to \emph{merge} these edges into one while preserving all shortest paths.
		The resulting single edge $e''$ from $u$ to $v$ gets the TTF $\min(f,f')$ defined by $\tau\mapsto\min\{f(\tau),f'(\tau)\}$.
		It can be computed in $\Oh{|f|+|f'|}$ time and $|\!\min(f,f')| \in \Oh{|f|+|f'|}$ holds.
\end{itemize}

In a time-dependent road network, shortest paths depend on the departure time.
For given start node $s$ and destination node $t$ there might be different shortest paths for different departure times.
The minimal travel times from $s$ to $t$ for all departure times $\tau$ are called the \emph{travel time profile} from $s$ to $t$ and are represented by a TTF.

\myparagraph{Approximations.}
Give a TTF $f$.
A \emph{lower bound} is a TTF $f^\downarrow$ with $f^\downarrow \le f$ and a \emph{lower $\varepsilon$-bound} if further $(1-\varepsilon)f \le f^\downarrow$.
An \emph{upper bound} is a TTF $f^\uparrow$ with $f \le f^\uparrow$ and an \emph{upper $\varepsilon$-bound} if further $f^\uparrow \le (1+\varepsilon)f$.
An \emph{$\varepsilon$-approximation} is a TTF $\fapprox$ with $(1-\varepsilon)f \le f^\updownarrow \le (1+\varepsilon)f$.
Approximate TTFs usually have fewer points and are therefore faster to process and require less memory.
To compute $\varepsilon$-bounds and $\varepsilon$-approximations from an exact TTF $f$ we use the efficient geometric algorithm described by Imai and Iri~\cite{ii-a-87}.
It yields a TTF with minimal number of points for $\varepsilon$ in time $\Oh{|f|}$.

\subsection{Time-Dependent Profile Dijkstra}
To compute the travel time profile from a source node $s$ to all other nodes, we can use a label correcting modification of Dijkstra's algorithm \cite{or-spmda-90}.
The modifications are as follows:
\begin{itemize}
\item \emph{Node labels.}
Each node $v$ has a tentative TTF from $s$ to $v$.
\item \emph{Priority queue (PQ).}
The keys used are the global minima of the labels.
Reinserts into the PQ are possible and happen (\emph{label correcting}).
\item \emph{Edge Relaxation}. Consider the relaxation of an edge $(u,v)$ with TTF $f_{uv}$.
Let the label of node $u$ be the TTF $f_u$.
The label $f_v$ of the node $v$ is updated by computing the minimum TTF of $f_v$ and $f_{uv}*f_u$.
\end{itemize}

% %%%%%%%%%%%%%%%%%%%%%%%%%%%%%%%%%%%%%%%%%%%%%%%%%%%%%%%%%%%%%%%%%%%%%%%%%%%%%
\subsection{Time-Dependent Contraction Hierarchies}
% %%%%%%%%%%%%%%%%%%%%%%%%%%%%%%%%%%%%%%%%%%%%%%%%%%%%%%%%%%%%%%%%%%%%%%%%%%%%%

\myparagraph{Hierarchies.}
In a \emph{time-dependent contraction hierarchy}~\cite{bdsv-tdch-09} all nodes of $G$ are \emph{ordered} by increasing `importance' \cite{gssd-chfsh-08}.
In order to simplify matters, we identify each node with its importance level, i.e. $V=1..n$.

Now, the TCH is constructed by \emph{contracting} the nodes in the above order.
Contracting a node $v$ means removing $v$ from the graph without changing shortest path distances between the remaining (more important) nodes.
This way we construct the next higher level of the hierarchy from the current one.
A trivial way to contract a node $v$ is to introduce a shortcut edge $(u, w)$ with TTF $g*f$ for every path $u \to_f v \to_g w$ with $v < u,w$.
But in order to keep the graph sparse, we can try to avoid a shortcut $(u, w)$ by finding a \emph{witness} -- a travel time profile $W$ from $u$ to $v$ fulfilling $W \leq g*f$.
Such a witness proves that the shortcut is never needed.
The node ordering and the construction of the TCH are performed offline in a precomputation and are only required once per graph independent of $S$ and $T$.

\myparagraph{Queries.}
In the point-to-point scenario, we compute the travel time profile between source $s$ and target $t$ by performing a bidirectional time-dependent profile search in the TCH.
The special restriction on a TCH search is that it only goes \emph{upward}, i.e. we only relax edges where the other node is more important.
This property is reflected in the \emph{upward graph} $\Gup \Is (V,\Eup) \mbox{ with } \Eup \Is \set{(u,v) \in E \mid u < v}$ and, the \emph{downward graph} $\Gdown  \Is (V,\Edown)$ with $\Edown \Is \set{ (u,v) \in E \mid u > v}$.
Both search scopes meet at \emph{candidate} nodes $u$ giving lower/upper bounds on the travel time between source and target, allowing us to prune the following profile search.
The bidirectional profile search computes forward TTF $f_u$ and backward TTF $g_u$ representing a TTF $g_u*f_u$ from source to target (though not necessarily an optimal one).
The travel time profile is $\min\set{g_u*f_u \mid u\text{ candidate}}$.

\section{Our Algorithm}

Given a source node $s$, we want to compute for each node $u$ in the graph the travel time profile $\td(u)$ from $s$ to $u$.
We initialize $\td(s) := (\tau \mapsto 0)$ and all other TTFs $\td(u) = (\tau \mapsto \infty)$.
Then, we perform a forward search from $s$ in $\Gup$ and update the tentative travel time profile $\td(u)$ for all nodes visited by this search.
Now, we process all nodes in the graph by descending importance level and compute
\begin{equation}
\td(u) := \min\big(\td(u), \min\set{ \link{\td(v)}{f_v} \mid v \to_{f_v} u \in \Edown }\big)
\label{eq:min}
\end{equation}

So essentially, we scan through all incoming downward edges $v \to_f u$ of $u$, link them to $\td(v)$ and build the minimum.
Doing this naively is very costly, as the travel time functions are complex.
As not all of the nodes $v$ contribute to the minimum in the end, we improve the performance by using pruning techniques developed for the time-dependent travel time table computation.
Assume that we have lower/upper $\varepsilon$-bounds $\lb{f_v}$ / $\ub{f_v}$ for $f_v$ and $\lb{\td(v)}$ / $\ub{\td(v)}$ for $\td(v)$.
Then we can use Algorithm~\ref{algo:mxn:td:all:min} to accelerate the computation of $\td(u)$.

\begin{algorithm}[H]
\caption{BuildMinimum($u$)}
\label{algo:mxn:td:all:min}
$\overline{\td} := \min_{v \to_{f_v} u \in \Edown}\set{\max f_v+\max\td(v)}$\tcp*[r]{upper bound based on maxima}\label{algo:mxn:td:all:min:tdmin}
$(\underline{v} \to \cdot) := \operatorname*{argmin}_{v \to_{f_v} u \in \Edown}\set{\min f+\min \td(v)}$\tcp*[r]{minimum node}\label{algo:mxn:td:all:min:cmin}
$\td^\uparrow := \link{\ub{\td(\underline{v})}}{\ub{f_{\underline{v}}}}$\tcp*[r]{upper bound based on approximate TTFs}\label{algo:mxn:td:all:min:tdminbegin}
$\overline{\td} := \min(\overline{\td},\max\td^\uparrow)$\tcp*[r]{tighten upper bound}
\ForEach(\tcp*[f]{loop over all downward edges}){$v \to_{f_v} u \in \Edown$}{
  \If(\tcp*[f]{prune using minima}){$\min f_v  +\min\lb{\td(v)} \le \overline{\td}$}{\label{algo:mxn:td:all:min:tdminprune}
    $\ub{\td} := \min\left(\td^\uparrow, \link{\ub{\td(v)}}{\ub{f_v}}\right)$\tcp*[r]{update upper bound}\label{algo:mxn:td:all:min:tdminend}
  }
}
$\td(u) := \min\left(\td(u), \link{\td(\underline{v})}{f_{\underline{v}}}\right)$\tcp*[r]{tentative travel time profile}\label{algo:mxn:td:all:min:tdbegin}
\ForEach(\tcp*[f]{loop over all downward edges}){$v \to_{f_v} u \in \Edown$}{
  \If(\tcp*[f]{prune using lower bounds}){$\neg(\link{\lb{\td(u)}}{\lb{f_v}} > \td^\uparrow)$}{\label{algo:mxn:td:all:min:tdprune}
    $\td(u) := \min\left(\td(u), \link{\td(v)}{f_v}\right)$\tcp*[r]{update travel time profile}\label{algo:mxn:td:all:min:eval}\label{algo:mxn:td:all:min:tdend}
  }
}
\end{algorithm}

We pass three times through the incoming downward edges $v\to_{f_v} u \in \Edown$.
\begin{enumerate}
 \item In Line~\ref{algo:mxn:td:all:min:tdmin} we compute an upper bound $\overline{\td}$ based on the maxima of $f_v$ and $\td(v)$.
  Also, in Line~\ref{algo:mxn:td:all:min:cmin} we compute the edge with minimum sum of the minima of $f_v$ and $\td(v)$.
  This edge is usually very important and a good starting point to obtain an tight lower bound.
 \item In Lines~\ref{algo:mxn:td:all:min:tdminbegin}--\ref{algo:mxn:td:all:min:tdminend} we compute an upper bound $\ub{\td}$ based on the upper $\varepsilon$-bounds.
  This bound is tighter than the one based on the maxima.
 \item In Lines \ref{algo:mxn:td:all:min:tdbegin}--\ref{algo:mxn:td:all:min:tdend} we compute the travel time profile and use the upper bound $\ub{\td}$ for pruning.
  So we only execute the very expensive link and minimum operations on $f_v$ and $\td(v)$ at Line~\ref{algo:mxn:td:all:min:eval}.
\end{enumerate}

In comparison, a Profile Dijkstra would update $\td(u)$ gradually when he processes node $v$, and thus cannot pass through all the downward edges several times to compute the intermediate upper bounds $\overline{\td}$ and $\ub{\td}$.

\section{Improvements}
Some of the improvements from \cite{dgnw-phast-10} can be applied, especially the reordering of nodes, and the parallel processing.
However, we did not adopt SIMD instructions or GPU, as we now operate on complex TTFs.

\section{Core-based Computation}
An interesting observation is that \cite{dgnw-phast-10} cannot be used like Dijkstra's algorithm to compute the distances to close-by nodes, as the nodes are no longer processed in order of increasing distance.
However, we can compute the distances only for a core of the contraction hierarchy, that is a number of $k$ most important nodes.

Computing only the distances to all core nodes is faster, especially in the time-dependent scenario.
Also, it takes much less space, as the TTFs usually contain thousands of points.

An application of the core-based computation is the computation of arc flags \cite[\S7.2]{dgnw-phast-10}, but now only for a core.
In \cite{bdsssw-chgds-10} it is observed that this brings significant speed-ups for the static scenario.
However, in the time-dependent scenario, upper/lower bounds used to compute exact arc flags, as using exact TTFs is too time-consuming \cite{d-tdsr-09}.
These arc flags are very weak, and simple heuristic computation of the arc flags using time-sampling significantly accelerate the query, but provide no approximation guarantee \cite{d-tdsr-09}.
So we expect that when we are able to compute exact arc flags with exact TTFs, this provides both a fast and exact query.

% %%%%%%%%%%%%%%%%%%%%%%%%%%%%%%%%%%%%%%%%%%%%%%%%%%%%%%%%%%%%%%%%%%%%%%%%%%%%%
\section{Experiments}
% %%%%%%%%%%%%%%%%%%%%%%%%%%%%%%%%%%%%%%%%%%%%%%%%%%%%%%%%%%%%%%%%%%%%%%%%%%%%%

\myparagraph{Input.}
We use a real-world time-dependent road network of Germany with 4.7 million nodes and 10.8 million edges, provided by PTV AG for scientific use.
It reflects the midweek (Tuesday till Thursday) traffic collected from historical data, i.e., a high traffic scenario with about 8 \% time dependent edges.
%We verify the robustness of our algorithms on other inputs in Appendix~\ref{section:appendix:other_inputs}.

\myparagraph{Hardware/Software.}
%The experiments were done on a machine with two Intel Xeon X5550 processors (Quad-Core) clocked at 2.67 GHz with 48 GiB of RAM and 2x8 MiB of Cache running SUSE Linux 11.1.
The experiments were done on a machine\footnote{The machine used in \cite{bgns-tdcha-10,gs-etdmm-10} is currently repaired, we plan to publish results for this machine later.} with two Intel Xeon E5345 processors (Quad-Core) clocked at 2.33 GHz with 16 GiB of RAM and 2x8 MiB of Cache running SUSE Linux 11.1.
We used the GCC 4.3.2 compiler with optimization level~3.

\myparagraph{Basic setup.}
We use a preprocessed TCH as input file \cite{bdsv-tdch-09}.
We use a core-size of 10\,000, and select 100 core nodes uniformly at random as source of our query.

The experimental results are presented in Table~\ref{table:query}.
The number of threads corresponds to the number used to answer a single query.

\begin{table}[t]
\centering
\setlength{\tabcolsep}{0.5\tabcolsep}
\begin{tabular}{c|rr|r}
algorithm & threads & prune $\varepsilon$ [\%]& query time [s] \\
\hline
Dijkstra & - & - & 116 \\
\hline
TCH & 1 & - & 105 \\
TCH & 1 & 10 & 108 \\
TCH & 1 & 1 & 48.2 \\
TCH & 1 & 0.1 & 32.6 \\
TCH & 1 & 0.01 & 35.6 \\
\hline
TCH & 2 & 0.1 & 16.0 \\
TCH & 4 & 0.1 & 8.8 \\
TCH & 6 & 0.1 & 6.5 \\
TCH & 8 & 0.1 & 5.7 \\
\end{tabular}
\caption{Performance.}
\label{table:query}
\end{table}

\section{Conclusion}
We presented an efficient algorithm for time-dependent one-to-all computation.
By applying it to a core, it can be used to accelerate precomputation of speed-up techniques.

\bibliographystyle{plain}
\bibliography{references}

\end{document}